\documentclass[aps,twocolumn,amsmath,amssymb,floatfix]{revtex4}
\usepackage{graphicx}
\usepackage{amssymb}
\usepackage{color}
\usepackage{float}
\usepackage{mathrsfs}
\usepackage{epstopdf}
\usepackage{placeins}
\usepackage[normalem]{ulem}  % \sout{old text} for strikeout

\renewcommand\sout{\bgroup \color{red} \ULdepth=-.5ex \ULset}
\begin{document}

%%%%%%%%%%%%%%%%%%%%%%%%%%%%%%%%%%%%%%%%%%%%%%%%%%%%%%%%%%%%%%%%%%%%%%%%%%%

%\begin{frontmatter}

%\title{Competition between fusion and quasifission processes in heavy ion collisions close to the Coulomb barrier}
\title{Connecting the nuclear EoS to the interplay between fusion and quasifission processes in low-energy nuclear reactions}
%heavy ion collisions close to the Coulomb barrier}

\author{H. Zheng$^{1,2}$\footnote{Email address: zhengh@snnu.edu.cn}, S. Burrello$^{2,3}$,  M. Colonna$^{2}$\footnote{Email address: colonna@lns.infn.it}, D. Lacroix$^{4}$, G. Scamps$^{5}$}

\affiliation{$^{1}$ School of Physics and Information Technology, Shaanxi Normal University, Xi'an 710119,  China}
\affiliation{$^{2}$ Laboratori Nazionali del Sud, INFN, I-95123 Catania, Italy}
\affiliation{$^{3}$ Departamento de FAMN, Universidad de Sevilla, Apartado 1065, E-41080 Sevilla, Spain}
\affiliation{$^{4}$ Institut de Physique Nucl\'eaire, IN2P3-CNRS, Universit\'e Paris-Sud, Universit\'e Paris-Saclay, F-91406 Orsay Cedex, France}
%\affiliation{$^{4}$ Dipartimento di Fisica, Universit\`a degli Studi di Milano, Via Celoria 16, I-20133 Milano, Italy}
%\affiliation{$^{5}$ INFN, Sezione di Milano, Via Celoria 16, I-20133 Milano, Italy}
\affiliation{$^{5}$ Center for Computational Sciences, University of Tsukuba, Tsukuba 305-8571, Japan}

%%%%%%%%%%%%%%%%%%%%%%%%%%%%%%%%%%%%%%%%%%%%%%%%%%%%%%%%%%%%%%%%%%%

\begin{abstract}
Within the Time Dependent Hartree Fock (TDHF) approach, we investigate the impact of several ingredients of the nuclear effective interaction, such as incompressibility, symmetry energy, effective mass, derivative of the Lane potential and surface terms on the exit channel (fusion vs quasifission) observed in the reaction $^{238}$U+$^{40}$Ca, close to the Coulomb barrier. Our results show that all the ingredients listed above contribute to the competition between fusion and quasifission processes, however the leading role in determining the outcome of the reaction is played by incompressibility, symmetry energy and the isoscalar coefficient of the surface term. This study unravels the complexity of the fusion and quasifission reaction dynamics and helps to understand the microscopic processes responsible for the final outcome of low energy heavy ion collisions in terms of relevant features of the nuclear effective interaction and associated equation of state (EoS). %, such as incompressibility and symmetry energy.

\end{abstract}

\pacs{}

\maketitle

\section{Introduction}
Understanding the dissipation mechanisms occurring in low energy heavy ion collisions represents one of the most challenging problems in nuclear reaction and structure studies \cite{Sch84,Fel84,Lac04,Sim10,Lac14,Lac15,Nak16}. 
Crucial information is provided by the investigation of strongly damped collisions of nuclei, that may lead to (incomplete) fusion, quasifission or deep-inelastic processes,  looking at the degree of equilibration reached along the reaction path and at the features of the final reaction products \cite{Baran:2004ih, Zheng:2016vio, Zheng:2016jrf, Rizzo:2010px, Tanimura:2016gnv, Tanimura:2015cua, Williams:2017egm, Hinde:2018vqp,Umar:2016vzo}. In addition, in these reactions one can observe effects reflecting a delicate interplay between the microscopic single-particle dynamics and the possible occurrence of collective motion \cite{Zheng:2016vio, Zheng:2016jrf, Rizzo:2010px, Savran:2013bha,Sim07}. 

In particular, dissipative reaction dynamics plays an essential role in the synthesis of superheavy elements (SHE), a quite appealing challenge of modern nuclear physics \cite{Hinde:2018vqp, Lazarev:1995zz, Oganessian:1999zza,Oganessian:2010zz, Oganessian:2012zz, Oganessian:2012ju, Oganessian:2015vpp, Khuyagbaatar:2014pxa, Khuyagbaatar:2015hjj}. The synthesizing process, realized by fusing two heavy nuclei in the laboratory,  can be schematically divided into three steps where both nuclear structure and dynamics are important: 1) the two nuclei find each other and their surfaces stick together; 2) the shape of the two nuclei evolves to form a compound nucleus; 3) the evaporation residue survives against statistical fission decay. These processes are usually related to the capture, compound nucleus formation and survival probabilities (see for instance \cite{Ada00,Ada04}) and two experimental methods, i.e., cold fusion with target nuclei close to the doubly magic nucleus $^{208}$Pb and hot fusion with actinide target nuclei \cite{Hinde:2018vqp, Umar:2016vzo}, have been devised so far.  
%The successful formation of superheavy nuclei is a complicated balance among the three steps, which results in two methods in experiments, i.e., cold fusion with target nuclei close to the doubly magic nucleus $^{208}$Pb and hot fusion with actinide target nuclei \cite{Hinde:2018vqp, Umar:2016vzo}. 
It is rather clear that, apart from the occurrence of statistical fission, the probability to form superheavy elements is strongly affected by the competition between compound nucleus formation and  quasifission processes. 

At low energies (close and/or above the Coulomb barrier), heavy ion reactions are governed, to a large extent, by one-body dissipation mechanisms (see for instance \cite{Fro96}). From a microscopic point of view, their description can be addressed within the Time Dependent Hartree Fock (TDHF) approach \cite{Lac02,Was08,Ayi09,Was09,Yil11,Yil14,Tanimura:2016gnv, Tanimura:2015cua, Williams:2017egm, Hinde:2018vqp, Umar:2016vzo, Vo-Phuoc:2016mgh, Oberacker:2014zsa,Simenel:2012xq, Wakhle:2014fsa, Umar:2010wf},  and its semi-classical approximation (the Vlasov equation) \cite{Baran:2004ih, Zheng:2016vio, Zheng:2016jrf, Rizzo:2010px}.
%\com{[Stefano: what about the collisions? Do we need to comment about this?]} 
These mean-field approaches (including stochastic extensions, that account for quantum
fluctuations) provide a suitable framework to study the many-body system at a  fully microscopic level and have been successfully applied to describe fusion reactions, nucleon transfer and deep-inelastic collisions, as well as the quasifission dynamics \cite{Simenel:2010my, Simenel:2010ty,Scamps:2012iv, Sekizawa:2013tga, Wakhle:2014fsa, Williams:2017egm}. Among possible examples, the prominent role of one-body dissipation is corroborated by the consistent results obtained, between TDHF and Vlasov calculations,  in studies related to the Giant Dipole Resonance (GDR) and for charge asymmetric nuclear reactions just above the Coulomb barrier \cite{Zheng:2016vio,burrello18prc},  and by the excellent quantitative agreement between recent quasifission experimental results and model calculations which incorporate one-body dissipation and fluctuations \cite{Williams:2017egm}. In spite of the apparent simplicity of the reaction dynamics, quite intriguing features may manifest along the fusion/fission path, reflecting the complexity of the self-consistent mean-field. Indeed, apart from the expected sensitivity to the properties, such as charge, mass and deformation,  of the two colliding nuclei, the reaction path is quite influenced also by the ingredients of the nuclear effective interaction employed in the calculations.  It should be noticed that the latter is closely connected to the nuclear Equation of State (EoS), which plays an important role in nuclear  structure \cite{Scamps:2013sca, Scamps:2014ica}, dynamics of heavy ion collisions at intermediate energy \cite{Baran:2004ih, Zheng:2016vio, Zheng:2016jrf, Rizzo:2010px,Chomaz:2003dz, Bonasera:1994zz} and astrophysical phenomena as well \cite{Giuliani:2013kna, Li:2008gp, Lattimer:2006xb}.
 
By considering collisions between either neutron poor or neutron rich systems,  the impact of the isospin degree of freedom on the reaction dynamics
%fusion cross section, on the competition between fusion and quasifission processes and on the charge equilibration rate reached in deep-inelastic collisions 
has been explored in theoretical studies \cite{Baran:2004ih, Zheng:2016vio, Zheng:2016jrf, Rizzo:2010px, Tanimura:2016gnv, Tanimura:2015cua, Williams:2017egm, Hinde:2018vqp, Umar:2016vzo, Vo-Phuoc:2016mgh, Oberacker:2014zsa,Simenel:2012xq, Wakhle:2014fsa, Umar:2010wf}. 
%\com{[Stefano: not clear for me! The sentence before this comment.]}  
Several investigations have also been devoted to the role of specific ingredients of the nuclear effective interaction.  %on the reaction dynamics.  
In particular, the 
influence of the symmetry energy, that is closely linked to the neutron-skin thickness, on the amplitude of the sub-barrier fusion cross section of neutron-rich nuclei has been evidenced in Ref. \cite{Reinhard:2016sce}. Other studies were dedicated to the sensitivity of isospin equilibration to the effective interaction, either in low-energy reactions, where collective pre-equilibrium dipole oscillations take place \cite{Zheng:2016vio, Zheng:2016jrf, Rizzo:2010px, baranPRL2001, pierrPRC2009, tson2001}, or for reactions at Fermi energies, where a sizable pre-equilibrium nucleon emission is observed \cite{Baran:2004ih, Baran:2005, Li:2008gp, Zhang:2015xna}. Reactions close to the Coulomb barrier,
%concerning in particular 
and more specifically 
at the frontier between fusion and other channels are also known to be sensitive to %specific terms of the effective interaction like 
the spin orbit term \cite{Uma86,Mar06,Dai14} or the tensor interactions \cite{Ste16,Luo18}. 

In this paper, we aim at getting a deeper understanding of the interplay between fusion and quasifission processes in low energy heavy ion collisions. As stressed above, this is particularly important in the search of new SHE. In keeping with the spirit of previous studies \cite{Zheng:2016jrf, Reinhard:2016sce, Zhang:2015xna}, we investigate, within the TDHF approach, the impact of relevant ingredients of the nuclear effective interaction, such as incompressibility, symmetry energy, effective mass, Lane potential derivative, and surface terms, 
%\com{[Do we need to change the order of these terms?]} 
on the exit channel (fusion vs quasifission) of central 
heavy ion reactions close to the Coulomb barrier. 
%%% agito qui
%the reaction $^{238}$U+$^{40}$Ca at $E_{cm}=203$ MeV and impact parameter b=0 fm. This reaction has been investigated in 
%great detail, for a specified EoS and in the TDHF framework,  
%%by Simenel and his collaborators  
%in previous papers \cite{Oberacker:2014zsa, Umar:2015lxa}, that we will consider 
%as a reference for our study. 
%%These studies can help us to choose the reasonable collision energy to make the contact time of the two colliding n%uclei not so long when quasifission happens, which makes our study possible. 
%It should be noticed that $^{40}$Ca is spherical and this will reduce the number of collision configurations and the complexity of the calculations \cite{Umar:2010wf, Wakhle:2014fsa}. On the other hand, effects linked to the deformation of the target will also be discussed.  
Our goal is to establish possible connections between the reaction dynamics and global nuclear matter properties, 
%such as incompressibility and symmetry energy, 
in density regions around and below the saturation value.  Such a comprehensive study
%, taking into account several ingredients of the effective interaction, 
is important to reach more reliable predictions about the probability to get compound nucleus formation. In turn, the comparison with available experimental data would allow one to extract information on specific aspects of the nuclear interaction, which are still poorly known. %\com{[Stefano: I would like to delete the last two sentences starting from on the one hand.]}

The paper is organized as follows. In section II, we introduce the theoretical framework as well as the set of  EoS employed in the calculations. In particular, we will consider EoS only differing by one ingredient, with respect to a reference case,  to focus on the effect of that particular ingredient on the reaction process. In such a way, we can decouple the correlations among the different sectors of the EoS. In section III, we present the results 
obtained for selected reactions, discussing how the several EoS ingredients affect the exit channel.  Conclusions and perspectives are drawn in section IV.

\section{Theoretical framework and effective interactions}
In the TDHF theory, the evolution of the one-body
density matrix $\hat {\rho}(t)$ is determined by, 
\begin{equation}
i\hbar \partial_{t}{\hat \rho}(t)=[h[{\hat \rho}],{\hat \rho}(t)], 
\label{EQ:TDHF}
\end{equation}
where $h[{\hat \rho}]={\bf p}^{2}/2m+U[{\bf p},{\rho}]$ is the mean-field Hamiltonian with $U$ as the self-consistent potential and $\rho({\bf r})$ denoting 
the local density. Within the Density Functional Theory, 
the starting point is the 
%one may start considering a given 
energy density functional $\mathscr{E}[\rho]$, from which 
the corresponding nuclear EoS and the potential $U$ can be %extracted
consistently derived.  
%from a 
%considered energy density functional, denoted by $\mathscr{E}(\rho)$, from which 
%and one may also 
%derive consistenly the corresponding nuclear EoS. 
%where $\rho$ is the local density of nuclear matter. In TDHF approach, within the quantum framework, $\rho$, still determined by the EoS, represents the associated one-body density matrix and the dynamical evolution of the system is
%\begin{equation}
%i\hbar \frac{d\rho}{dt}=[h[\rho], \rho],
%\end{equation}
%where $h[\rho]$ is the self-consistent mean field. 

In the present work, 
we adopt Skyrme effective interactions, which are characterized in terms 
of nine interaction parameters ($t_0, t_1, t_2, t_3, x_0, x_1, x_2, x_3, \sigma$), plus the spin-orbit coupling constants $W_{0(i)}$ \cite{Skyrme,Vau72,Ben03,Sto07}.
%
%T.H.R. Skyrme, Philos. Mag. 1, 1043 (1956); Nucl. Phys. 9, 615 (1959); 9, 635 (1959).
Apart from the spin-orbit term, the energy density is expressed in terms of the isoscalar, $\rho=\rho_n+\rho_p$, and isovector, $\rho_{3}=\rho_n-\rho_p$,  densities and kinetic energy densities ($\tau=\tau_{n}+\tau_{p}, \tau_{3}=\tau_{n}-\tau_{p}$) as \cite{radutaEJPA2014, Zheng:2016vio, lipp89, Lesinski:2006cu}:
\begin{eqnarray}
\mathscr{E}[\rho]&\equiv& \mathscr{E}_{kin}(\tau) + \mathscr{E}_{pot}(\rho,\rho_3, \tau, \tau_3) \nonumber \\
&=& \frac{\hbar^2}{2 m}\tau + C_0\rho^2 + D_0\rho_{3}^2 + C_3\rho^{\sigma + 2} + D_3\rho^{\sigma}\rho_{3}^2 ~ \nonumber \\
&+& C_{eff}\rho\tau + D_{eff}\rho_{3}\tau_{3} \nonumber \\
&+& C_{surf}(\bigtriangledown\rho)^2 + D_{surf}(\bigtriangledown\rho_3)^2,
\label{eq:rhoE}
\end{eqnarray}
where the coefficients $C_{..}$, $D_{..}$ are combinations of the standard Skyrme parameters (see Appendix \ref{app:coeff}).
In particular, the terms with coefficients $C_{eff}$ and $D_{eff}$ are the momentum dependent contributions to the nuclear 
effective interaction.  The Coulomb interaction is also considered in the calculations.  
%\textcolor{blue}{Denis: the next sentence is not clear to me? This seems to be an easy general statement?}
It turns out to be useful to explicit the relations between the coefficients of the 
Skyrme interaction and relevant nuclear properties. 
%In Refs.\cite{Chen:2009wv, Chen:2010qx} one can find the relations connecting the Skyrme parameters to relevant nuclear properties:
In analogy with the studies of Refs. \cite{Chen:2009wv, Chen:2010qx}, we will consider: 
saturation density $\rho_0$; energy per nucleon of symmetric nuclear matter at $\rho_0$ ($E_0$); incompressibility $K_0$; isoscalar effective mass $m_{s}^*$ and isovector effective 
mass $m_{v}^*$ at saturation density; symmetry energy at $\rho_0$ ($J$); slope of the symmetry energy at $\rho_0$ ($L$); strength of the isoscalar 
surface term $G_S = C_{surf}/2$ and  strength of the isovector surface term $G_V = -D_{surf}/2$.
 
By this connection, it becomes straightforward to explore the impact of specific 
%ingredients of the nuclear effective interaction 
nuclear matter properties on the reaction dynamics. 
% in the reaction $^{238}$U+$^{40}$Ca, close to the Coulomb barrier by varying one of the macroscopic observables in MSL and obtaining the corresponding parameters for standard Skyrme EoS in TDHF.
Here, instead of the isovector effective mass, we prefer to employ the derivative, with 
respect to the momentum ${p}$, of the Lane potential, which has a more intuitive physical
meaning, related to the splitting of neutron and proton effective masses, $m_n^*$ and $m_p^*$. 
%The single particle potential is
Denoting by $U_i=\displaystyle \left. \frac{\partial\mathscr{E}_{pot}}{\partial \rho_i}\right|_{p}$ the single
particle potential, where $i$ stands for neutrons or protons and $\mathscr{E}_{pot}$ is the
potential part of the energy density functional, Eq.(\ref{eq:rhoE}),   
the Lane potential is written as
\begin{eqnarray}
U_{Lane}&=& \frac{U_n-U_p}{2I}\nonumber\\
&=&2D_0\rho+2D_3\rho^{\sigma+1}+\frac{D_{eff}\tau_3}{I}+D_{eff}\rho\frac{p^2}{\hbar^2}, \label{lanep}
\end{eqnarray}
where $I=\frac{\rho_n-\rho_p}{\rho}$ is the asymmetry parameter. 
Therefore, the Lane potential derivative reads:
\begin{equation}
\frac{dU_{Lane}}{dp}=\frac{2D_{eff}}{\hbar^2}\rho p = \left(\frac{1}{m_s^*}-\frac{1}{m_v^*}\right)p = 
\frac{f_I}{m}p,\label{dudp}
\end{equation}
where %$m_s^*$ and $m_v^*$ denote isoscalar and isovector effective masses and 
the parameter  $f_I$ has been introduced and $m$ denotes the nucleon mass. 
$f_I$ actually gives a measure of the neutron-proton effective mass splitting because the 
following relation holds: 
$\frac{1}{m_s^*}-\frac{1}{m_v^*} = \frac{1}{2I}
\left(\frac{1}{m_n^*}-\frac{1}{m_p^*}\right)$ \cite{Zhang:2015xna}.

In our study, we will consider the recently introduced SAMi-J Skyrme effective interactions \cite{coll4}. The corresponding parameters have been determined based on the SAMi fitting protocol \cite{coll4}: binding energies and charge radii of some doubly magic nuclei, which allow the SAMi-J family to predict a reasonable saturation density, energy per nucleon and incompressibility modulus of symmetric nuclear matter; %see table \ref{table1}; 
some selected spin-isospin sensitive Landau-Migdal parameters \cite{caoPRC2010};  the neutron matter EoS of Ref. \cite{wiringaPRC1988}. According to the strength of the momentum dependent terms, these interactions lead to an effective isoscalar nucleon mass $m^*_{s} = 0.675~m$ and a neutron-proton effective mass splitting $m^*_n - m^*_p = 0.023~mI$ MeV at saturation density, with the corresponding parameter $f_I=-0.0251$. This small mass 
splitting effect is associated with a quite flat momentum dependence of the symmetry potential.  
It should be noticed that the SAMi-J interactions 
%are characterized by different values of the symmetry energy slope $L$ and 
exhibit
a correlation between $J$ and $L$, so that all interactions lead to the same value of the symmetry energy below normal
density (at $\rho \approx 0.6\rho_0$), well describing the ground state properties of nuclei.
For neutron-rich nuclei, the interactions with a larger $L$ (and $J$) value predict a thicker neutron skin (see Tables \ref{ca40sami} and \ref{U238sami}).

%\end{document}

\section{Results and discussions}
We have performed TDHF calculations for the system $^{238}$U+$^{40}$Ca, at $E_{cm}=203$ MeV and 
zero impact parameter.   % \cite{Oberacker:2014zsa, Umar:2015lxa}. 
 This reaction has been investigated in 
great detail, for a specified EoS and in the TDHF framework,  
%%by Simenel and his collaborators  
in previous papers \cite{Oberacker:2014zsa, Umar:2015lxa}, that we will consider 
as a reference for our study. 

In practice, we use the EV8 code to initialize the two nuclei \cite{Bon05} and the TDHF-3D code developed in Refs. \cite{Kim97,Lac02,Was08,Scamps:2012iv}
to follow the reaction dynamics. 
% \textcolor{red}{(Denis: please complete by providing some more details on the dynamics about mesh step, time step, total mesh size...)}
We adopt a 3D lattice mesh ($96\times 40\times 20$) with a mesh step of 0.8 
fm, and a time step $\Delta t = 0.36$ fm/c. The initial distance between the two colliding nuclei is 26.4 fm.
% for the TDHF-3D simulation with time step $\Delta t = 0.36$ fm/c.} \textcolor{red}{(Denis: please complete by providing some more details on the dynamics about mesh step, time step, total mesh size...)}
 %%%HERE. 
%%These studies can help us to choose the reasonable collision energy to make the contact time of the two colliding n%uclei not so long when quasifission happens, which makes our study possible. 

The values of binding energy, neutron and proton root mean square radii 
and quadrupole deformation parameter are reported in Tables \ref{ca40sami} and \ref{U238sami},
%\ref{ca40sami}, 
%\ref{U238sami}, 
for the three SAMi-J interactions that we will consider in
our analysis.  

\begin{table}[htbp]
\begin{center}
\begin{tabular}{|c|c|c|c|c|c|c|}
\hline
Interaction  & $\sqrt{\langle r^2 \rangle_n}$ & $\sqrt{\langle r^2 \rangle_p}$  & neutron skin & $\beta_2$   & BE/A (MeV) \\
\hline
SAMi-J27 & 3.360 & 3.410 & -0.050  & 0.0  & -8.210  \\
\hline
SAMi-J31 & 3.357 & 3.405  & -0.048 & 0.0 & -8.374  \\
\hline 
SAMi-J35 & 3.350 & 3.396 & -0.046  & 0.0  &  -8.507 \\
\hline
\end{tabular}
\caption{Neutron and proton root mean square radii, and their difference,  quadrupole deformation
and binding energy  for $^{40}$Ca, as obtained  with the SAMi-J interactions.
The experimental value of the binding energy is BE/A=-8.551 MeV \cite{jag, nndc}.} \label{ca40sami}
\end{center}
\end{table}

\begin{table}[htbp]
\begin{center}
\begin{tabular}{|c|c|c|c|c|c|c|}
\hline
Interaction  & $\sqrt{\langle r^2 \rangle_n}$ & $\sqrt{\langle r^2 \rangle_p}$  & neutron skin & $\beta_2$   & BE/A (MeV) \\
\hline
SAMi-J27 & 5.927 & 5.802 & 0.125 & 0.251 & -7.547  \\
\hline
SAMi-J31 & 6.02 & 5.81 & 0.21 & 0.228 & -7.524  \\
\hline
SAMi-J35 & 6.10 & 5.815 & 0.285 & 0.228 & -7.499  \\
\hline
\end{tabular}
\caption{Neutron and proton root mean square radii, and their difference, quadrupole deformation and
binding energy for $^{238}$U, as obtained  with the SAMi-J interactions.
The experimental value for the binding energy is BE/A=-7.570 MeV \cite{jag, nndc}.} \label{U238sami}
\end{center}
\end{table}

%%%%%%%%%%%%%%%%%%%%%%%%%%%%%%%%%%%%%%%%%%%%%%%%%%%%%%%%%%%%%%%%%%%%%%%%%%%%%%

\begin{figure}
\centering
\begin{tabular}{c}
\includegraphics*[scale=0.55]{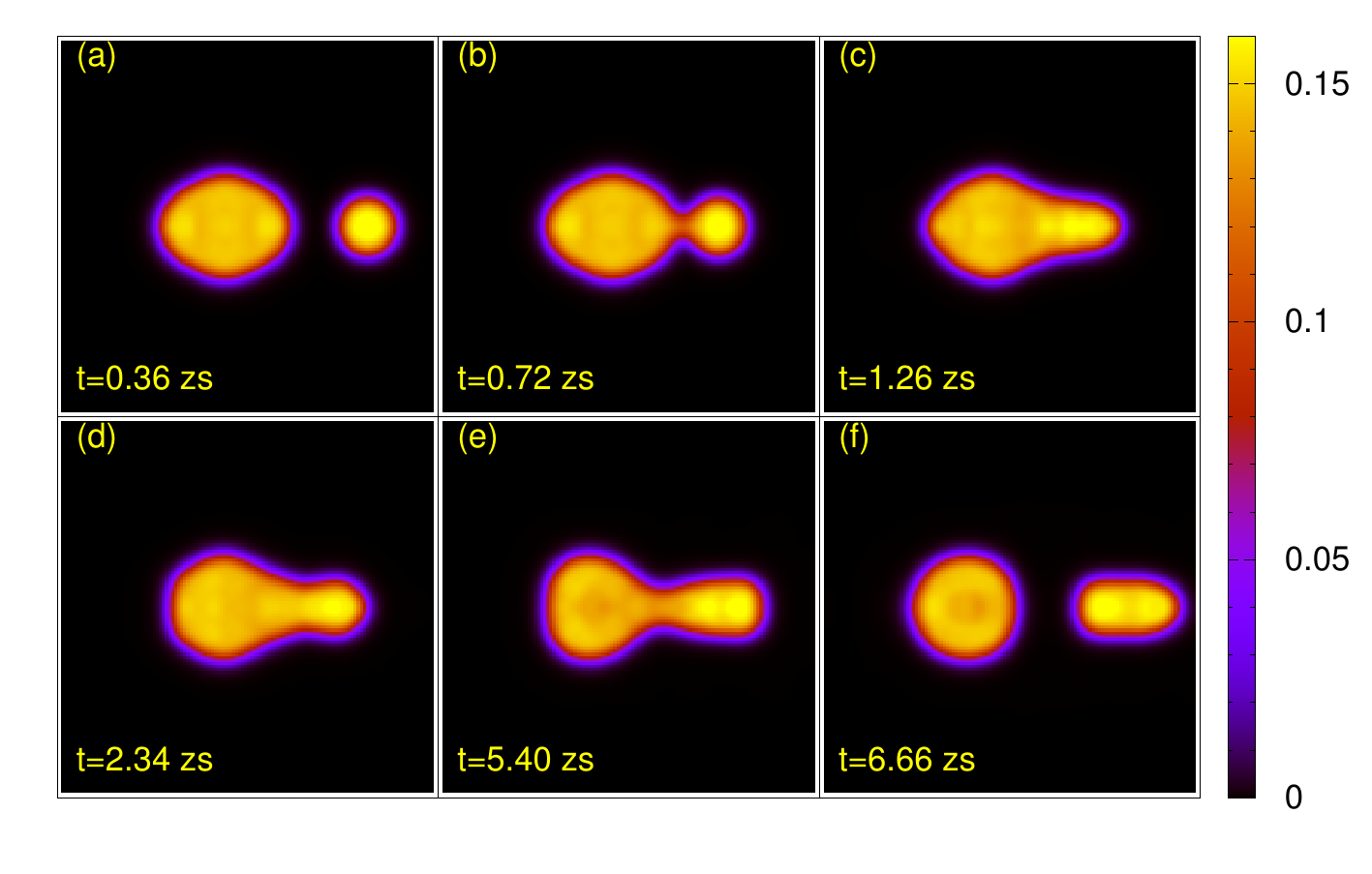}
\end{tabular}
\caption{(Color online) Density contour plot at different time instants for the tip collision of the reaction $^{238}$U+$^{40}$Ca (tip orientation) at $E_{cm}=203$ MeV and b=0 fm. The SAMi-J31 interaction is employed.}
\label{tipsm31}
\end{figure}

The beam energy considered is in the range of the
transition from fusion to quasifission processes, thus it is well adapted to our study 
of the competition between the two reaction mechanisms. 
It should be noticed that $^{40}$Ca is spherical and this will reduce the number of collision configurations and the 
complexity of the calculations \cite{Umar:2010wf, Wakhle:2014fsa}. 
On the other hand, %effects linked to the deformation of the target will also be discussed.  
since the ground state of $^{238}$U is deformed, it is worthwhile to consider reaction configurations
corresponding to two possible projectile-target orientations: side and tip.   
%\textcolor{green}{(Denis: I really feel that 
%the fusion barrier height (dynamical or in the frozen approximation) for the reference interaction would be 
%a valuable information--most probably referee will ask for it)}  
%\textcolor{red}{(Denis: I think it would be 
%meaningful to provide a table with the value of the deformation parameter $\beta_2$ in the Uranium for all interactions that are discussed in
%the text. If this changes much with interaction we might give some conclusion because we already know how the fusion barrier depends 
%on $\beta_2$)}
For the tip orientation, at the energy considered, quasifission is always observed 
in TDHF calculations \cite{Oberacker:2014zsa, Simenel:2012xq, Wakhle:2014fsa}. We 
will show just one tip collision case, as an example, and then concentrate on
side collisions in our study. 
The trajectory of the reaction is traced by evaluating 
the quadrupole moment $Q_2(t)=\langle 2x^2-y^2-z^2\rangle$ of the composite system, 
with $x$ denoting the beam axis. An increasing trend of $Q_2(t)$ indicates that the
system is evolving towards quasifission. On the other hand, if $Q_2(t)$ stays around a constant value, then the system fuses. To check the outcome of the reaction, one can also look directly at density 
contour plots
at different time instants, as shown in Fig. \ref{tipsm31}
%, we show density contour plots, at different time instants, obtained 
for the tip collision of the reaction considered, %$^{238}$U+$^{40}$Ca at $E_{cm}=203$ MeV and b=0 fm, 
with the effective interaction SAMi-J31. One can see that the quasifission happens in a few zs which is consistent with the statements in Refs. \cite{Oberacker:2014zsa, Simenel:2012xq, Wakhle:2014fsa}.
 % on the reaction plane. 
%\textcolor{blue}{The contour plot is not on the reaction plane now because we adopt Scamps' suggestion, it is perpendicular with the reaction plane. Maybe it is called out of plane.}
%evolution of the reaction system versus time to check its status in a direct manner. 
 
%In Fig. \ref{tipsm31}, we show density contour plots, at different time instants, obtained 
%for the tip collision of the reaction considered, %$^{238}$U+$^{40}$Ca at $E_{cm}=203$ MeV and b=0 fm, 
%with the effective interaction SAMi-J31. 
%One can see that the quasifission happens in a few zs which is consistent with the statements in Refs. \cite{Oberacker:2014zsa, Simenel:2012xq, Wakhle:2014fsa}.

\begin{figure}
\centering
\begin{tabular}{c}
\includegraphics*[scale=0.3]{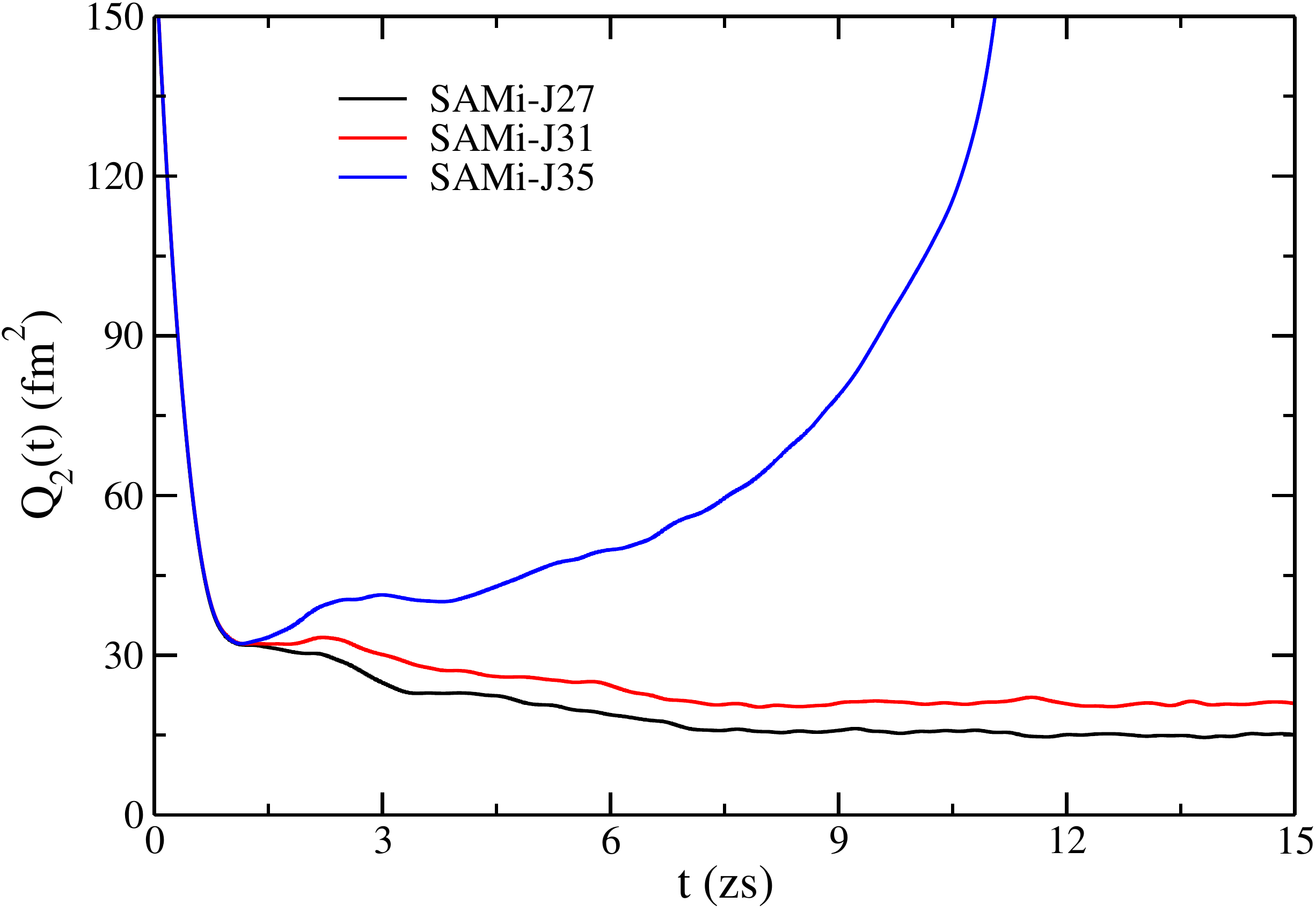}
\end{tabular}
\caption{(Color online) The quadrupole moment time evolution of the reaction $^{238}$U+$^{40}$Ca (side orientation) at $E_{cm}=203$ MeV and b=0 fm, for three SAMi-J EoS.}
\label{figuresm}
\end{figure}

\begin{figure}
\centering
\begin{tabular}{c}
\includegraphics*[scale=0.55]{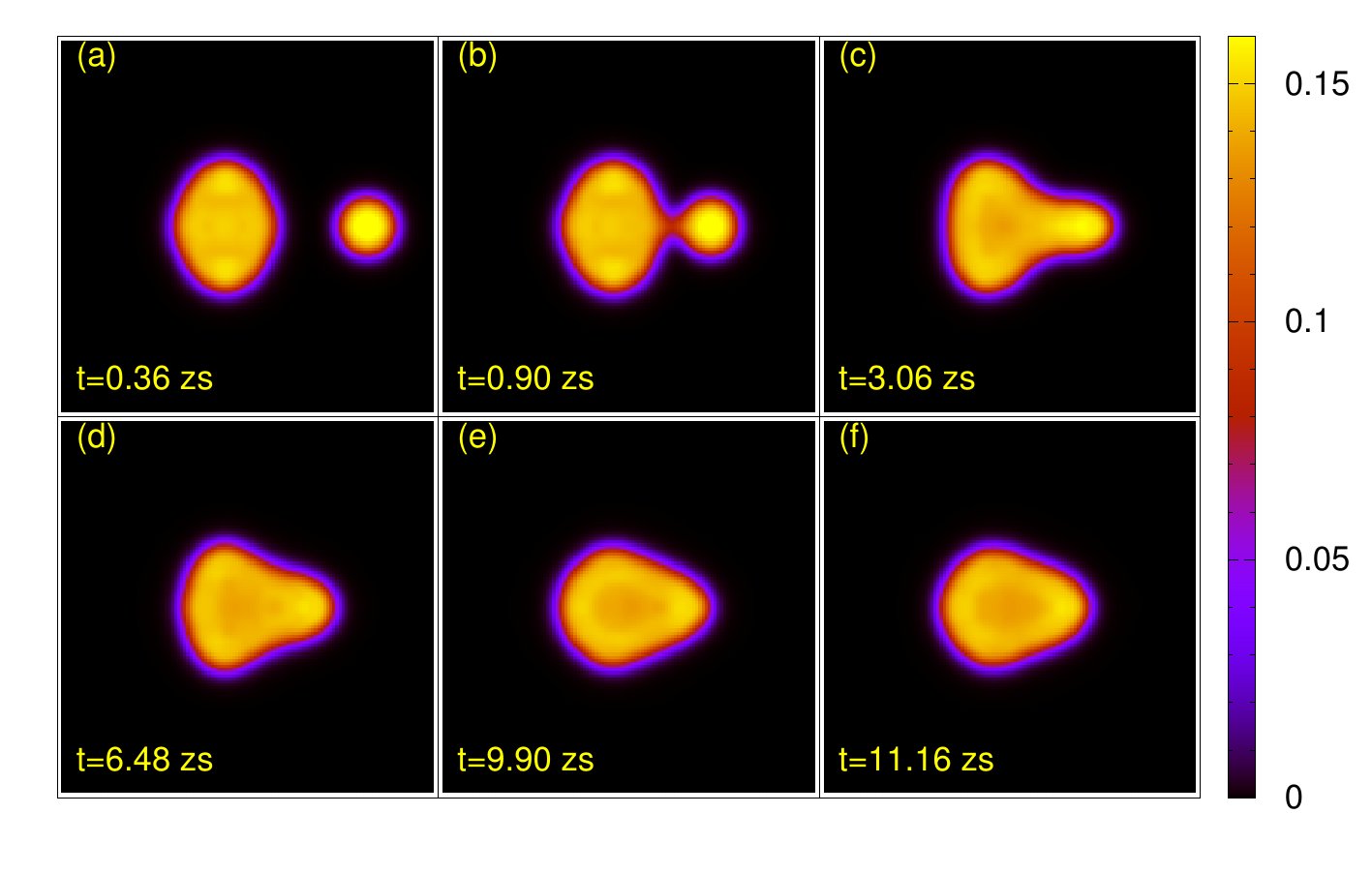}
\end{tabular}
\caption{(Color on line) Same as in Fig. \ref{tipsm31}, but for the side collision.}
\label{axialsm31}
\end{figure}

\begin{figure}
\centering
\begin{tabular}{c}
\includegraphics*[scale=0.55]{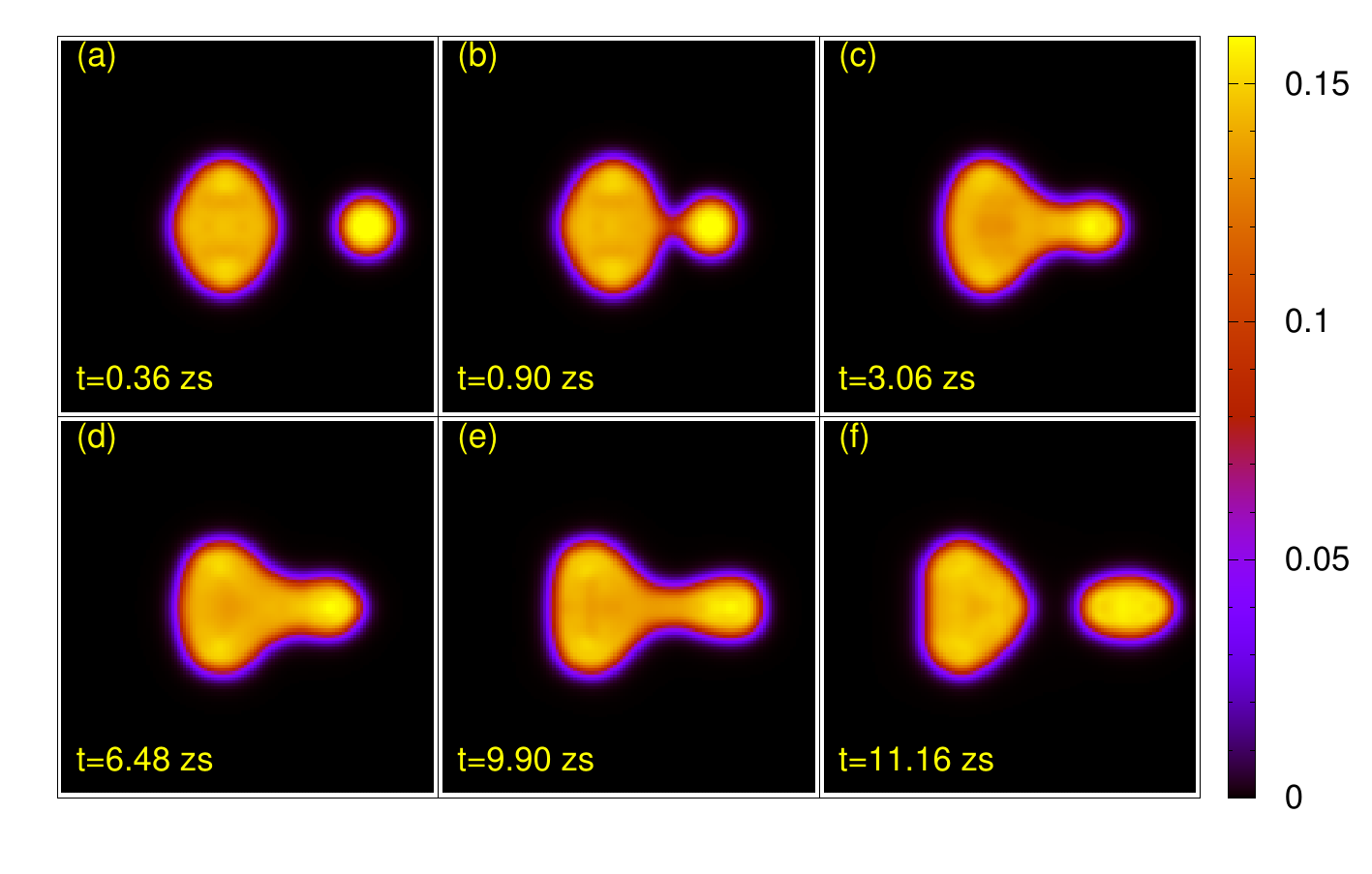}
\end{tabular}
\caption{(Color online) Same as in Fig. \ref{tipsm31}, but for the side collision, with SAMi-J35.}
\label{axialsm35}
\end{figure}

Let us now turn to discuss side collisions. In Fig. \ref{figuresm}, the time evolution of the quadrupole moment of the composite system is shown for calculations corresponding to 
three different SAMi-J interactions, namely SAMi-J27, SAMi-J31 and SAMi-J35, whose label denotes the symmetry 
energy value (J) at saturation density. 
Clearly, we can see that the SAMi-J35 parametrization leads to 
quasifission, whereas the other two SAMi-J EoS are associated with fusion. 
Corresponding density contour plots are shown in Figs. \ref{axialsm31} and \ref{axialsm35} for SAMi-J31 and SAMi-J35, respectively, in order to better display this dual behavior.  The competition between fusion and quasifission has been interpreted  
in terms of the features exhibited by the fusion barrier, $V_B$, as evaluated within the Density Constrained - TDHF method \cite{Umar:2016vzo, Umar:2010wf, Umar:2015lxa}, when employing different effective interactions. 
%But in this paper, we would like to study how the macroscopic observables affect on the exit channel (fusion vs. quasifission) in the low energy heavy ion collisions. 
The results shown on Fig. \ref{figuresm} would indicate that a larger symmetry energy slope, 
which results in a thicker neutron skin for neutron rich nuclei, makes the reaction system easier to separate.   However, as we will discuss in the following, the three 
parameterizations also differ by other aspects. 

To extend our discussion about the sensitivity of the reaction path to the ingredients
of the effective interaction, we will enlarge the set of Skyrme interactions employed
in the TDHF calculations. 
%But the three SAMi-J EoS have different surface terms which are very important for the low energy reaction intuitively. In the MSL model, the nine macroscopic observables are ($\rho_0, E_0(\rho_0), K_0, m_{s}^*, m_{v}^*, J, L, G_S, G_V$). Since we don't want to change $\rho_0$ and  $E_0(\rho_0)$ and the effect of the isovector coefficient of the surface term $G_V$ is small which we will show later, we have six macroscopic observables left in the effective interactions to study their impacts on the exit channel of low energy reaction. 
%For SAMi-J EoS, $J$ and $L$ are correlated in order to describe well the groud state properties of nuclei. Therefore we vary $J$ and $L$ at the same time to keep this merit when we study the impact of the symmetry energy in our study. 
%Then, 
In addition to the three SAMi-J parametrizations,  
we will consider interactions corresponding to the variations of six quantities, with respect to  the SAMi-J31, that is taken as a reference: the symmetry energy slope $L$ (keeping the $J-L$ correlation 
discussed above), the incompressibility, the effective mass, the parameter $f_I$ and the surface terms $G_S$ and $G_V$.
$\rho_0$ and $E_0$  are adopted from SAMi-J31 for all the EoS.  We have checked that the binding energy and the root mean square of proton and neutron radii of $^{238}$U and $^{40}$Ca are well preserved (within a few percent) under those operations, with the neutron skin thickness of  $^{238}$U being mainly determined by the
symmetry energy slope $L$.    
For the convenience of description, we list the properties of the different EoS in Table \ref{table1}. The EoS name follows the convention that we only label the terms which are different, with respect to the ingredients of the SAMi-J31 parametrization.

\begin{table*}[htbp]
\begin{center}
\begin{tabular}{|c|c|c|c|c|c|c|c|c|c|c|c|c|}
\hline
No & EoS &  $\rho_0$ (fm$^{-3}$) & $E_0$ (MeV) &$K_0$ (MeV) & $J$ (MeV) & $L$ (MeV) & $m_{s}^*/m$ & $m_{v}^*/m$ & $f_I$ & $G_S$ & $G_V$  & Result \\
\hline
 & SAMi-J27  & 0.160 & -15.93 & 245 & 27 & 30 & 0.675 & 0.664 & -0.0251 & 149.2 & -8.6 & Fusion \\
\hline
S1 & SAMi-J31  & 0.156 & -15.83 & 245 & 31 & 74 & 0.675 & 0.664 &-0.0251 & 140.9 & 3.1& Fusion \\
\hline
  & SAMi-J35  & 0.154 & -15.69 & 245 & 35 & 115 & 0.675 & 0.664 &-0.0251 & 131.1 & 15.4 & Fission \\
\hline
S2 & J27 & 0.156 & -15.83  & 245 & 27 & 30 & 0.675 & 0.664 & -0.0251 & 140.9 & 3.1 & Fusion \\
\hline
S3 & J35 & 0.156 & -15.83 & 245 & 35 & 115 & 0.675 & 0.664 & -0.0251 & 140.9 & 3.1 & Fusion \\
\hline
 & Gs35 & 0.156 & -15.83 & 245 & 31 & 74 & 0.675 & 0.664 & -0.0251 & 131.1 & 3.1 & Fission \\
\hline
 & J35\_Gs35 & 0.156 & -15.83 & 245 & 35 & 115 & 0.675 & 0.664 & -0.0251 & 131.1 & 3.1&  Fission \\
\hline
 & J35\_Gv35 & 0.156 & -15.83 & 245 & 35 & 115 & 0.675 & 0.664 & -0.0251 & 140.9 & 15.4 & Fusion \\
\hline
 & J35\_Gs35Gv35 & 0.156 & -15.83 & 245 & 35 & 115 & 0.675 & 0.664 & -0.0251 & 131.1 & 15.4 & Fission \\
\hline
S4 & K200 &  0.156 & -15.83 & 200 & 31 & 74 & 0.675 & 0.664 & -0.0251 & 140.9 & 3.1 & Fission \\
\hline
S5 & K290 &  0.156 & -15.83 & 290 & 31 & 74  & 0.675 & 0.664 & -0.0251 & 140.9 & 3.1& Fusion \\
\hline
 %  & Sm31\_Sm27\_Gs35Gv31\_ms067 &  0.156 & -15.83 & 245 &  27 & 30 & 0.675 & 0.664 & -0.0251 & 131.1 & 3.1 \\
%\hline
%   & Sm31\_Sm27\_Gs35Gv31\_ms085 &  0.156 & -15.83 & 245 &  27 & 30 & 0.85 & 0.832 & -0.0251 & 131.1 & 3.1 \\
%\hline
 %  & Sm31\_Sm27\_Gs35Gv31\_ms100 &  0.156 & -15.83 & 245 &  27 & 30 & 1.0 & 0.976 & -0.0251 & 131.1 & 3.1 \\
%\hline
 %  & Sm31\_Sm27\_Gs35Gv31\_fI020 &  0.156 & -15.83 & 245 &  27 & 30 & 0.675 & 0.781 & 0.20 & 131.1 & 3.1 \\
%\hline
%  & Sm31\_Sm27\_Gs35Gv31\_fIn024 &  0.156 & -15.83 & 245 &  27 & 30 & 0.675 & 0.581 & -0.24 & 131.1 & 3.1 \\
%\hline
S6 & ms085 & 0.156 & -15.83 & 245 & 31 & 74 & 0.85 & 0.832 & -0.0251 & 140.9 & 3.1 & Fusion\\
\hline
S7 & ms100 & 0.156 & -15.83 & 245 & 31 & 74 & 1.0 & 0.976 & -0.0251 & 140.9 & 3.1 & Fusion \\
\hline
& Gs35\_ms085 & 0.156 & -15.83 & 245 & 31 & 74 & 0.85 & 0.832 & -0.0251 & 131.1 & 3.1 & Fusion \\
\hline
 & Gs35\_ms100 & 0.156 & -15.83 & 245 & 31 & 74 & 1.0 & 0.976 & -0.0251 & 131.1 & 3.1 & Fusion \\
\hline
S8 & fI020 & 0.156 & -15.83 & 245 & 31 & 74 & 0.675 & 0.781 & 0.20 & 140.9 & 3.1 & Fusion \\
\hline
S9 & fIn024 & 0.156 & -15.83 & 245 & 31 & 74 & 0.675 & 0.581 & -0.24 & 140.9 & 3.1 & Fusion \\
\hline
 & Gs35\_fI020 & 0.156 & -15.83 & 245 & 31 & 74 & 0.675 & 0.781 & 0.2 & 131.1 & 3.1 & Fission \\
\hline
 & Gs35\_fIn024 & 0.156 & -15.83 & 245 & 31 & 74 & 0.675 & 0.581 & -0.24 & 131.1 & 3.1 & Fission\\
\hline
%\hline
%S6 & Sm31-ms085 & & &245 & 31 & 74.4 & 0.85 & 0.832 & -0.0251 & &\\
%\hline
%S7 & Sm31-ms100 & &  & 245 & 31 & 74.4 & 1.0 & 0.976 & -0.0251 & & \\
%\hline
%& J27\_Gs35 & 0.156 & -15.83 & 245 & 27 & 30 & 0.675 & 0.664 & -0.0251 & 131.1 & 3.1 \\
%\hline
\end{tabular}
\caption{The properties of the different EoS considered in the present work, 
{and the exit channel of the reaction}. 
The units of $G_S$ and $G_V$ are MeV$\cdot$fm$^5$. Associated values of standard Skyrme parameters $(t_i, x_i)$ are given in Appendix 
\ref{app:coeff}.} \label{table1}
\end{center}
\end{table*}

\subsection{Symmetry energy effects}
\begin{figure}
\centering
\begin{tabular}{c}
\includegraphics*[scale=0.3]{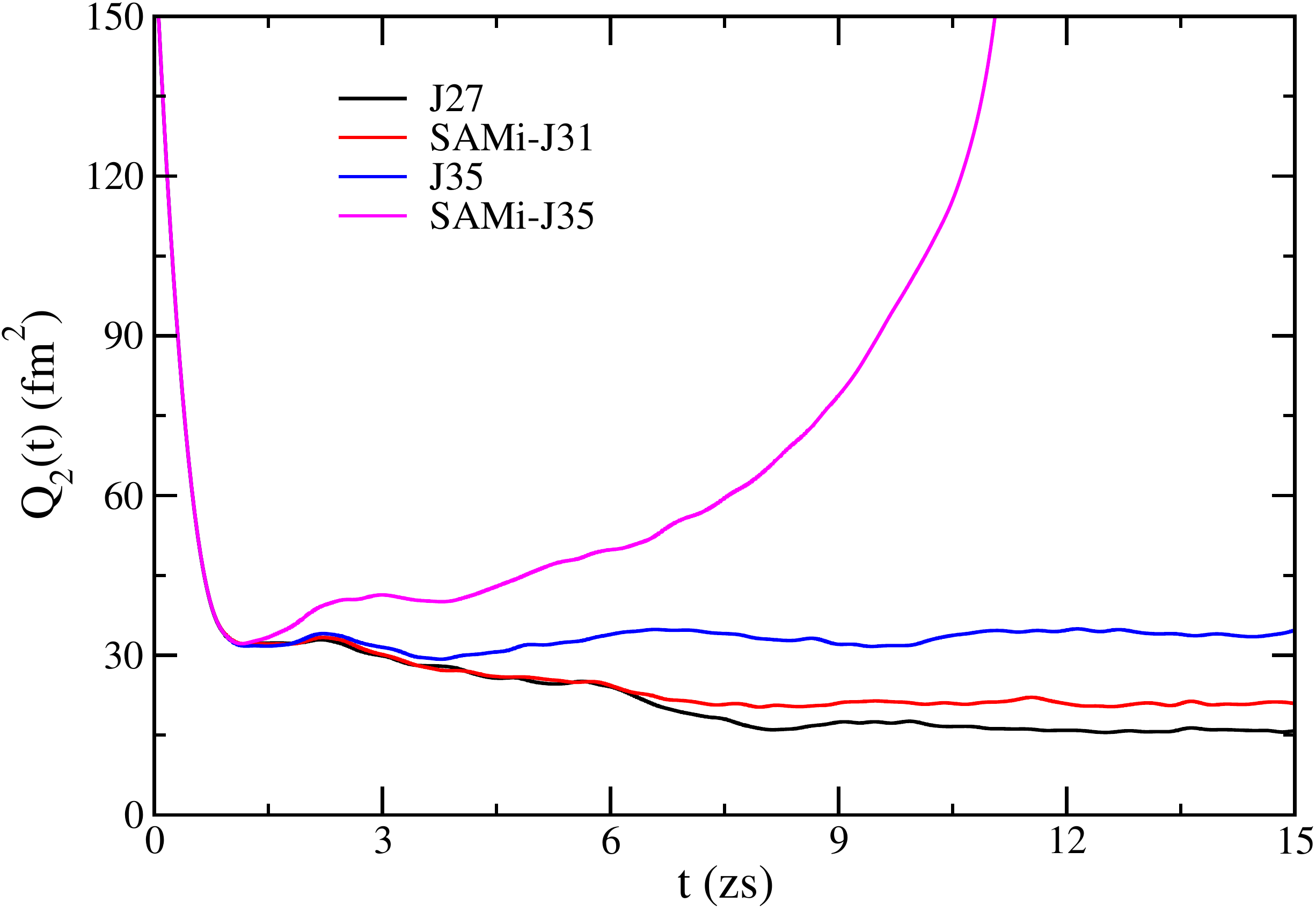}
\end{tabular}
\caption{(Color online) The symmetry energy dependence of the quadrupole moment evolution of the reaction considered 
(same as in Fig. \ref{figuresm}). 
All reactions are performed at $E_{cm}=203$ MeV and b=0 fm.}
\label{sym}
\end{figure}

In Fig. \ref{sym}, we show the quadrupole moment evolution of the reaction considered, 
for calculations employing the EoS obtained by varying the symmetry energy
properties (S1, S2 and S3).  One can notice that the result differs from what is shown in Fig. \ref{figuresm}, where the three SAMi-J parametrizations are compared. 
Indeed SAMi-J35 calculations (also reported in  Fig. \ref{sym}) lead to quasifission, where the S3 parametrization does not, though it presents the same symmetry energy features.  
This can be explained by considering that  
the three SAMi-J EoS are characterized not only by a different ($J$-$L$) combination, but also 
%$\rho_0$, $G_S$ and $G_V$, 
exhibit different surface properties, see Table \ref{table1}. 
%Here we create three new EoS using seven macroscopic observables same as in SAMi-J31 and adopting the $J$ and $L$ from the three SAMi-J EoS respectively. 
%As we 

However, one observes that the quadrupole moment associated with the EoS considered here is ordered by the symmetry energy, i.e., the larger the symmetry energy, the larger the quadrupole moment. 
Owing to the neutron excess in our system, a larger symmetry energy around normal density leads to a more repulsive dynamics, as one would intuitively expect.  
However, for the interactions considered, the quadrupole moment keeps quite smaller than the result associated with the SAMi-J35.  This indicates that surface terms may play a very important role in the reaction dynamics, as we will discuss in the following section.

\subsection{Surface term effects}

\begin{figure}
\centering
\begin{tabular}{cc}
\includegraphics*[scale=0.3]{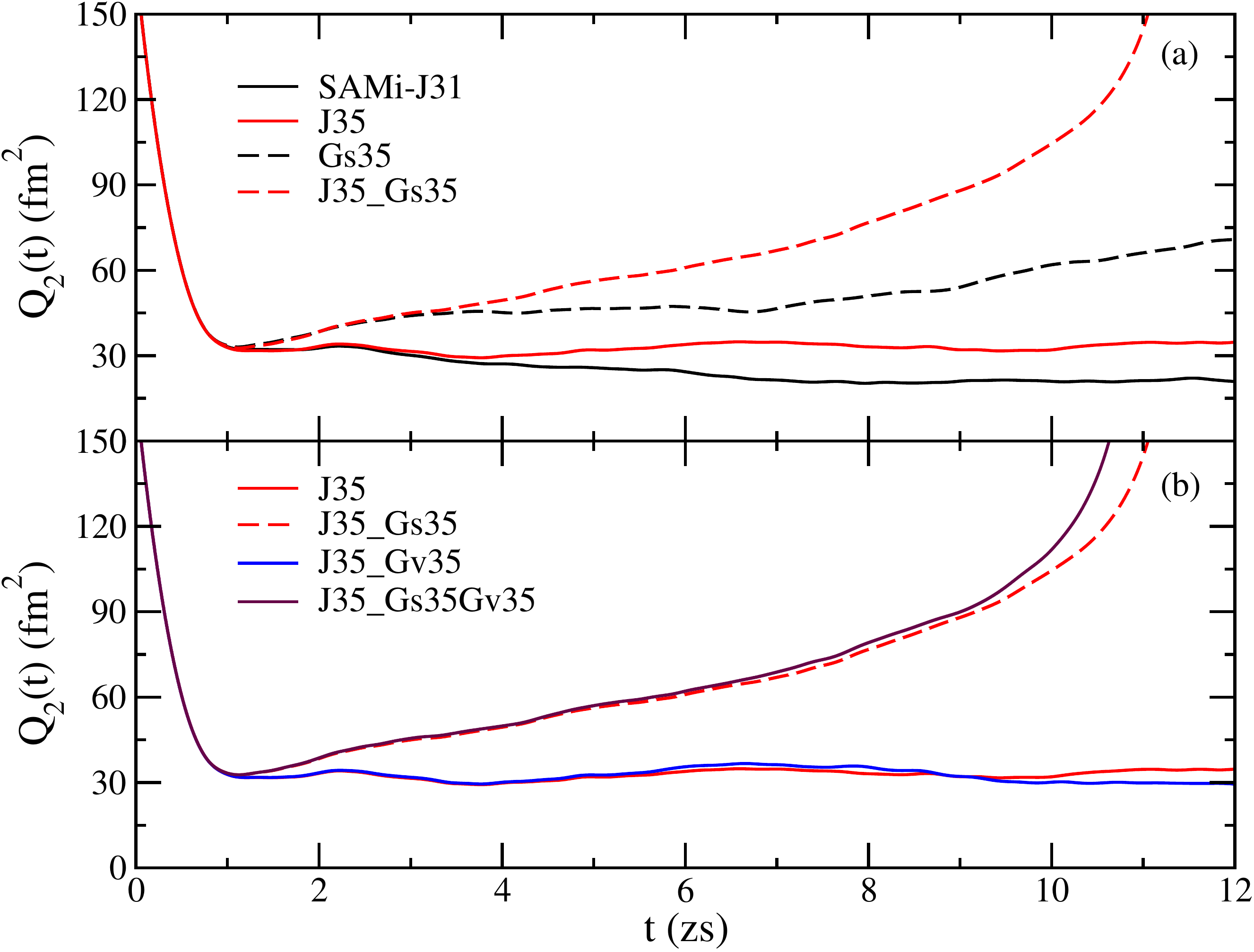}
\end{tabular}
\caption{(Color online) Surface effects on the quadrupole moment evolution of the reaction considered (same as in Fig. \ref{figuresm}). 
 All reactions are performed at $E_{cm}=203$ MeV and b=0 fm.}
\label{surf}
\end{figure}

In Fig. \ref{surf}, we show the time evolution of the quadrupole moment $Q_2$, as obtained for the 
EoS corresponding to different surface term ($G_S$ and $G_V$) combinations, as indicated in Table \ref{table1}. 
%starting from SAMi-J31 and J35. %and the $J$ and $L$ from SAMi-J35. 
As it is shown in panel (a), by comparing the results associated with SAMi-J31 and Gs35, a reduced $G_S$ surface term helps the system to separate, even in the case of a parametrization having
the same symmetry energy as SAMi-J31, that is not repulsive enough to lead to quasifission (see Fig. \ref{sym}).  
Combined with a larger symmetry energy (J35), the surface term reduction leads to a quite fast quasifission dynamics (J35\_Gs35). 
This result can be explained considering that, along the approaching phase a reduced surface term favors the formation of more elongated configurations, helping fission. 
On the other hand, by considering interactions where also $G_V$ is changed significantly (see Table \ref{table1}), panel (b) shows that the isovector surface term has only a tiny effect on the reaction dynamics.  
%The results show that the isoscalar coefficient of the surface term $G_S$ plays the dominant role to determine the average quadrupole evolution of the reaction system and the effect of the isovector coefficient of the surface term $G_V$ can be ignored. This is the reason we fixed the $G_V$ from SAMi-J31 in our previous investigations. In Fig. \ref{surf}(b), the average quadrupole evolution of three EoS with the same $J$ and $L$ as the three SAMi-J EoS and the isoscalar coefficient of the surface term $G_S$ from SAMi-J35 have been shown. The results of the three EoS in Fig. \ref{sym} with the isoscalar coefficient of the surface term $G_S$ from SAMi-J31 are also plotted for references. As we can see that the isoscalar coefficient of surface term determines the average quadrupole evolution of the reaction system at the very early stage, then the symmetry energy starts to play the role. 

\subsection{Effects of the incompressibility $K_0$}

\begin{figure}
\centering
\begin{tabular}{cc}
\includegraphics*[scale=0.3]{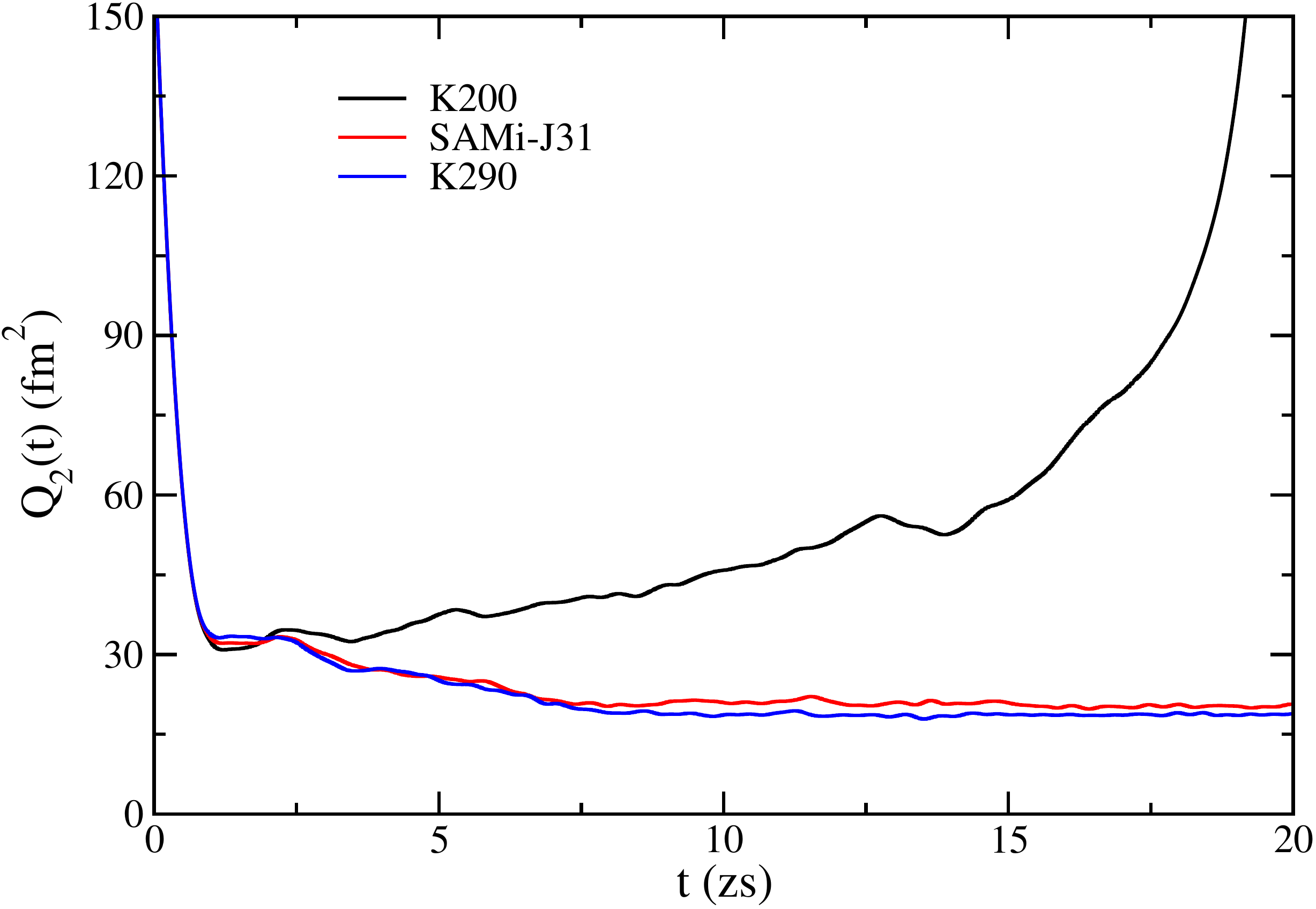}
\end{tabular}
\caption{(Color online) The incompressibility dependence of the quadrupole moment evolution, for the same
reaction as in Fig. \ref{figuresm}.  All reactions are performed at $E_{cm}=203$ MeV and b=0 fm.}
\label{kq}
\end{figure}

In Fig. \ref{kq}, we show the results corresponding to three EoS with different incompressibilities (S1, S4, S5). For $K_0$=200 MeV the system gets quasifission, whereas fusion is observed for the other two larger $K_0$ values. 
%whose energy per particle for symmetric infinite nuclear matter is higher than $K_0$=200 MeV beyond (below and above) the saturation density. 
The observed dependence  of our results on the incompressibility  
is related to the fact that it is more difficult to compress or to expand 
%when the density is above the saturation density and to expand when the density is below the saturation density 
the composite reaction system formed along the reaction path, if $K_0$ is large.
Indeed, in this case, the system needs to pay more energy to undergo density oscillations of
a given amplitude,  as compared to the calculations corresponding to smaller $K_0$ values. 
%Indeed, this is what we observe from the quadrupole moment evolution for different $K_0$ values. 
We observe that, at the compression stage, the system exhibits the smallest quadrupole moment for the case $K_0$=200 MeV, because the system is easier to compress and becomes more compact. %The opposite is for the case with the largest $K_0$. 
Then, along the expansion phase, the quadrupole moment increases significantly because
it is easier to expand the system towards 
densities below the saturation value. When the deformed system overcomes a given threshold, the reaction path will result in 
quasifission. Apparently, this does not happen for the cases corresponding to $K_0$=245 MeV (SAMi-J31) and $K_0$=290 MeV (S5), for which fusion is finally observed 
%do not overcome their thresholds and end up with fusions. At the same time, their average 
and the quadrupole moment oscillates around a constant value.
Clearly, density oscillations of larger amplitude help the system to fission. 

\subsection{Effects of the isoscalar effective mass $m_{s}^*$}

\begin{figure}
\centering
\begin{tabular}{cc}
\includegraphics*[scale=0.3]{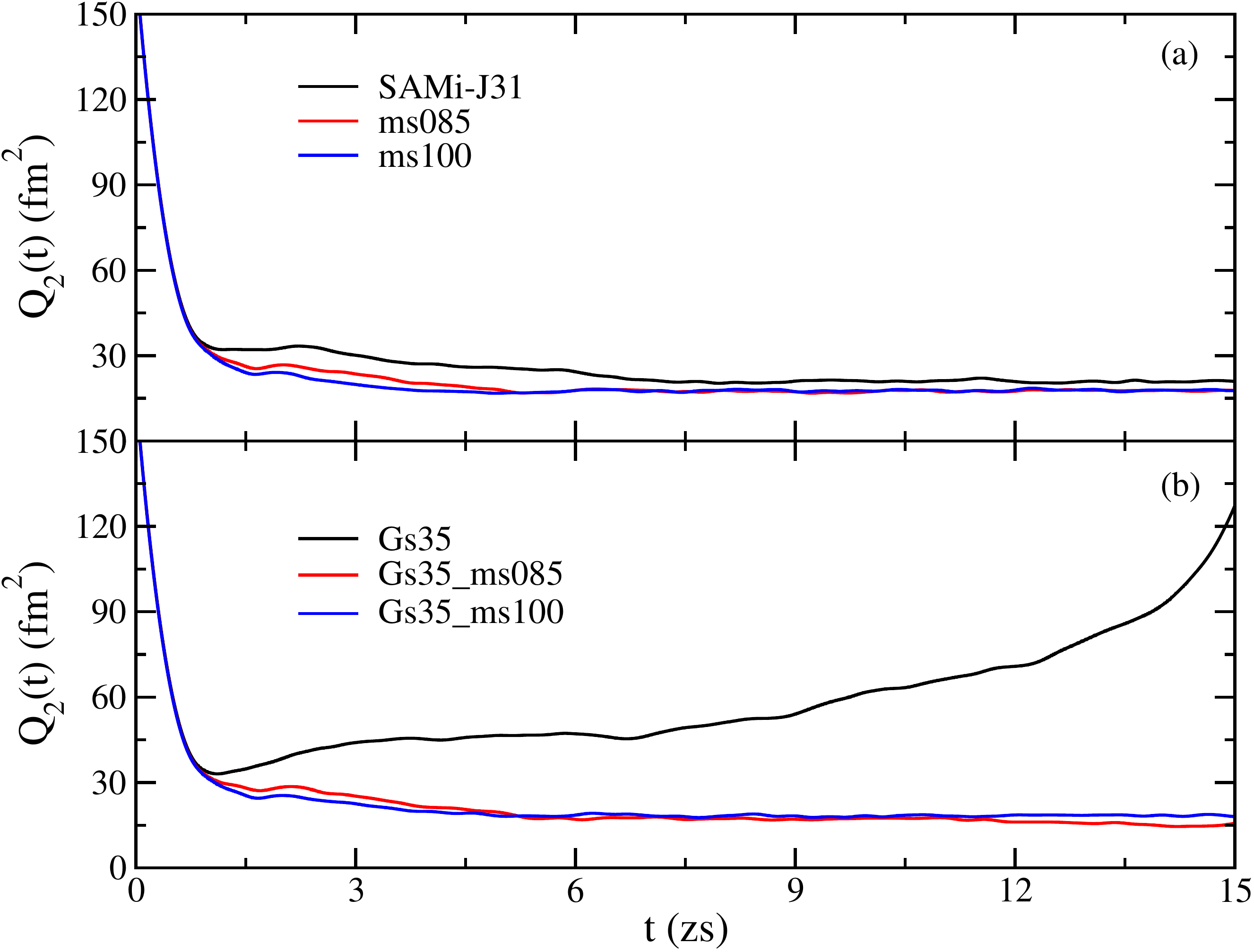}
\end{tabular}
\caption{(Color online) The dependence of the quadrupole moment evolution on the isoscalar effective mass.
Same reaction system as in Fig. \ref{figuresm}. All reactions are performed at $E_{cm}=203$ MeV and b=0 fm.}
\label{mq}
\end{figure}

The effect of the isoscalar effective mass on the quadrupole moment evolution 
of the reaction considered is shown in Fig. \ref{mq}, by considering parametrizations with larger
effective mass than the value associated with SAMi-J31 (S6, S7). 
One can see that, starting from a situation where fusion is observed (SAMi-J31), the
increase of the nucleon effective mass does not change the reaction dynamics; however the calculations corresponding to larger effective mass values lead to more compact configurations, associated with
a smaller quadrupole moment. 
%to the smallest isoscalar effective mass exhibits the largest quadrupole moment. 
In Fig. \ref{mq}(b), we explore the impact of the isoscalar effective mass also on a trajectory leading to quasifission (corresponding to the parametrization Gs35). 
In this case, it is observed that a larger effective mass changes the reaction dynamics, leading to fusion.
%the smallest isoscalar effective mass makes the system quasifission, whereas 
%the other two EoS, having larger isoscalar effective masses, lead to fusion. 
In the latter case, 
the quadrupole evolution exhibits the same pattern in panels (a) and (b). 
To understand these results, %associated with the smallest isoscalar effective mass, 
one may consider that particles having a smaller effective mass can move faster in the nuclear potential, so that  
for the system it is easier to escape from the attractive nuclear interaction and evolve towards quasifission.  
This is in line with what is observed in the study of collective modes (such as the GDR), where effective interactions with small effective mass lead to higher oscillation frequency and to a more abundant 
particle emission \cite{Zheng:2016vio}. 
%Therefore, they have the largest average quadrupole of the reaction system among the EoS adopted. 
One can also argue that particles with a small isoscalar effective mass can invert more easily their direction of motion, helping the system expansion. 
%When the reaction system is approaching each other, the case with smaller isoscalar effective mass can penetrate less deep than the larger isoscalar effective mass case. For the smaller isoscalar effective mass, 
On the other hand, a larger effective mass favors the trapping of the system into the nuclear potential, leading to fusion.
%the system can not overcome the fusion threshold and ends with quasifission in Fig. \ref{mq}(b) which is different from the counterpart in Fig. \ref{mq}(a). For the other two cases, they result in fusion. 
A careful inspection of Fig. \ref{mq}(b)  reveals that, at the early stage,  the quadrupole moment is larger in 
the case of the interaction with isoscalar effective mass equal to 0.85m, with respect to the `ms100' case ($m_s = m$), 
but this trend is inverted at a later stage. This can be attributed to the fact that momentum dependent interactions,  lead to a larger (smaller) repulsion 
for nucleons with  momenta larger (smaller) than the Fermi momentum, with respect to the `ms100' case.
 
Thus,
%We can use the same explanation for the smallest isoscalar effective mass case, at the beginning, the reaction system can not penetrate each other deeper than the case with larger isoscalar effective mass and it has the larger average quadrupole. 
once the system overcomes the fusion barrier,  more compact configurations are observed in the `ms085' case, indicating a larger attraction at small momenta. 
%the system is easier to reach equilibrium for the smaller isoscalar effective mass case because the particle is easier to change their motion status and it has the smaller average quadrupole.

\subsection{Effects of fI}

\begin{figure}[H]
\centering
\begin{tabular}{cc}
\includegraphics*[scale=0.3]{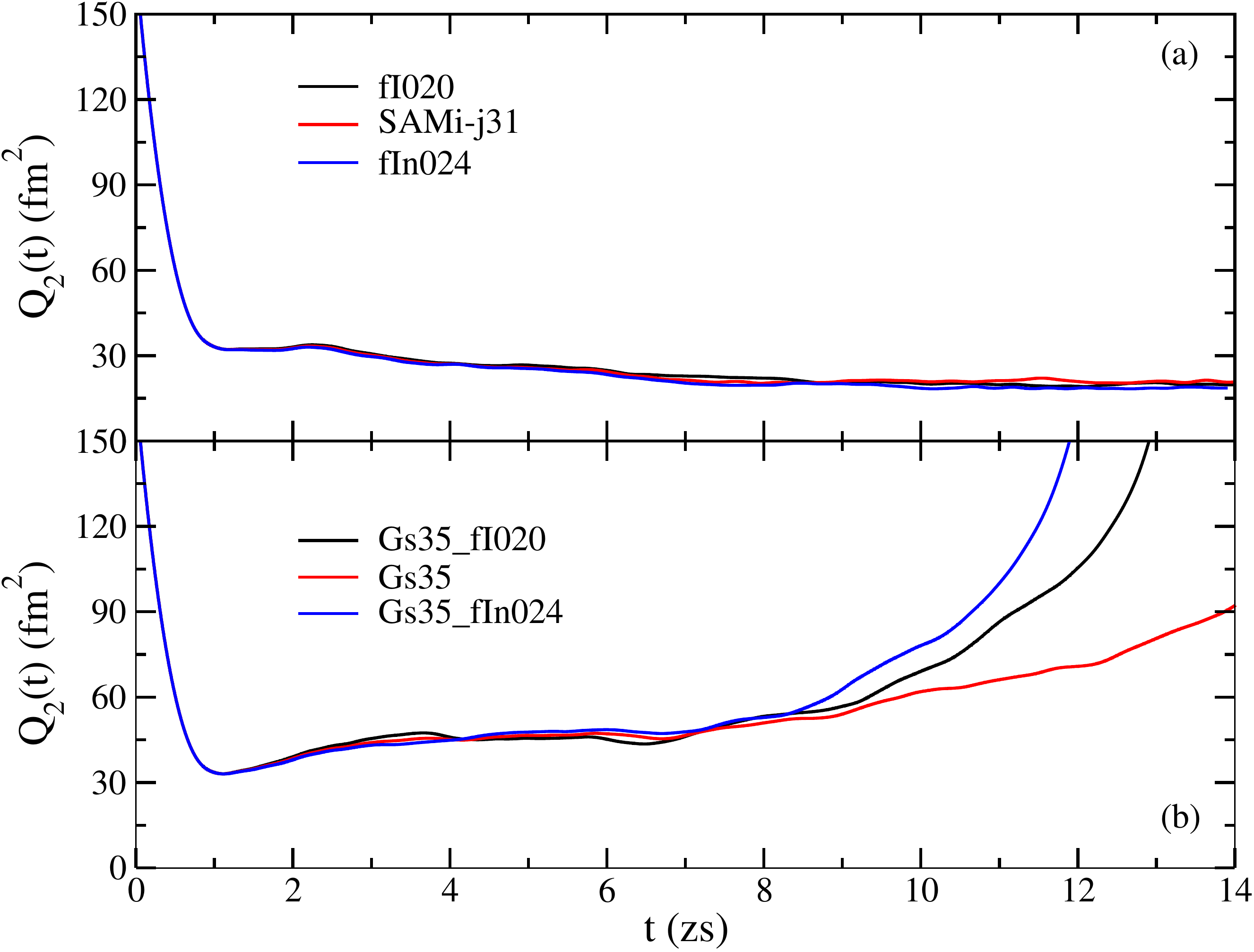}
\end{tabular}
\caption{(Color online) The $f_I$ dependence of the quadrupole moment evolution of the system considered (same as in Fig. \ref{figuresm}).  All reactions are performed at $E_{cm}=203$ MeV and b=0 fm.} 
\label{fi}
\end{figure}

We now concentrate on the reaction dynamics associated with 
%the quadrupole moment evolution for calculations corresponding to 
the three EoS having different values of $f_I$, adopting the isoscalar surface term of SAMi-J31 (S1, S8, S9), see Fig. \ref{fi}(a), and of SAMi-J35, see Fig. \ref{fi}(b).
One can observe that the system ends up with fusion for the three cases in panel (a) 
and quasifissions for the three cases in panel (b). 
Thus the $f_I$ parameter does not affect crucially the outcome of the reaction, either 
fusion or quasifission. 
The ordering observed in panel (b) may result from a delicate balance between symmetry energy, n/p effective
mass splitting and Coulomb repulsion effects.
Increasing the neutron-proton effective mass splitting, with $m^*_n < m^*_p$ (`fI020' case), leads to a larger
neutron repulsion, in addition to symmetry energy effects. As a result, we observe a faster quasifission dynamics. 
On the other hand, a n/p effective splitting of opposite sign,  with $m^*_p < m^*_n$ (`fIn024' case), 
tends to counterbalance symmetry energy effects. However, in this situation, the relative role of the
Coulomb repulsion is enhanced, that may also lead to a faster dynamics, as we actually observe in Fig. \ref{fi}(b).
It may be interesting to note that 
similar effects of the neutron/proton effective mass splitting are discussed
for observables typical of nuclear reactions in the Fermi energy domain, such as
the isotopic content of the pre-equilibrium 
nucleon emission \cite{Zhang:2015xna}.

%Even though the reaction systems end with quasifissions for the three cases, we can clearly see that the larger $f_I$ can accelerate the quasifission process. In the ground state of nuclei, the neutron potential is higher than the proton potential. When $f_I$ is positive, it makes the proton and neutron behavior more different which will result in more unstable of the reaction system. While $f_I$ is negative, it will reduce the difference between protons and neutrons and the reaction system becomes more stable comparing to the case $f_I>0$. In fact, we have seen the ordering of the quadrupole evolution for the three $f_I$. 

%{\color{blue}Do we need the figure with three EoS ending with fusions? The derivative of the lane potential is proportional to the momentum $p$. Therefore, it has the larger impact on the nucleons where the density is higher (bulk). $f_I$ can't determine the destine of the reaction system: fusion or quasifission because the surface term is dominant. But when the system will end with quasifission, $f_I$ will accelerate the fission process as I explain above. When the system will end with fusion, $f_I$ will only have small impact. It seems that we also can explain the results for the fusion cases.}

%\section{Theoretical framework}

%\begin{figure}
%\begin{center}
%\includegraphics*[scale=0.36]{transden_sn132_sami31_IS.eps}
%\end{center}
%\caption{{\color{blue}Similar to Fig.\ref{Trhosn132sami31IV} bur with IS initial perturbation.}}
%\label{Trhosn132sami31IS}
%\end{figure}

\begin{table*}[thbp]
\begin{center}
\begin{tabular}{|c|c|c|c|c|c|c|c|c|c|c|}
\hline
No & EoS &  $t_0$  & $t_1$  &$t_2$  & $t_3$ & $x_0$  & $x_1$ & $x_2$ & $x_3$ & $\sigma$ \\
\hline
  & SAMi-J27  & -1876.09 & 481.087 & -75.7069 & 10184.6 & 0.482235 & -0.557967 & 0.213066 & 1.00219 & 0.254634 \\
\hline
S1 & SAMi-J31  & -1844.28 & 460.727 & -110.200 & 10112.4 & -0.0237088 & -0.458608 & -0.431251 & 0.00764843 & 0.268372  \\
\hline
  & SAMi-J35  & -1799.53 & 436.229 & -144.972 & 9955.45 & -0.443908 & -0.343557 & -0.783861 & -0.882427 & 0.284323 \\
  \hline
  S2 & J27 &   -1844.27 &    460.727 &   -110.200 &  10112.4 &    0.478794 &   -0.458608 &   -0.431252 &    1.012559 &    0.268374  \\
\hline
S3 & J35  &  -1844.27 &    460.727 &   -110.200 &  10112.4 &   -0.461008 &   -0.458608 &   -0.431252 &   -0.879839 &    0.268374  \\
\hline
 & Gs35 &  -1844.28 &    434.803 &    -84.2766 &  10112.4 &   -0.0237087 &   -0.456140 &   -0.410106 &    0.00764882 &    0.268374  \\
 \hline
  & J35\_Gs35 &  -1844.27 &    434.803 &    -84.2767 &  10112.4 &   -0.461008 &   -0.456140 &   -0.410106 &   -0.879839 &    0.268374  \\
\hline
 & J35\_Gv35 &  -1844.27 &    460.727 &   -175.617 &  10112.4 &   -0.461008 &   -0.352118 &   -0.736234 &   -0.879839 &    0.268374  \\
\hline
 & J35\_Gs35Gv35 &  -1844.27 &    434.803 &   -149.694 &  10112.4 &   -0.461008 &   -0.343301 &   -0.777144 &   -0.879839 &    0.268374  \\
\hline
S4 & K200 &   5698.04 &    460.727 &   -110.200 & -36164.8 &    0.0177978 &   -0.458608 &   -0.431251 &    0.00764843 &   -0.0421665 \\
\hline
S5 & K290 &  -1295.07 &    460.727 &   -110.200 &   8342.72 &   -0.0370067 &   -0.458608 &   -0.431251 &    0.00764843 &    0.578726 \\
\hline
S6 & ms085 &  -1696.12 &    406.841 &   -271.859 &  11451.3 &   -0.105374 &   -0.453125 &   -0.472133 &   -0.281046 &    0.354121  \\
\hline
S7 & ms100 &  -1654.78 &    375.621 &   -365.519 &  12510.9 &   -0.130771 &   -0.449229 &   -0.479273 &   -0.397744 &    0.388782\\
\hline
& Gs35\_ms085 &  -1696.12 &    380.917 &   -245.935 &  11451.3 &   -0.105374 &   -0.449935 &   -0.469195 &   -0.281046 &    0.354121  \\
\hline
 & Gs35\_ms100 &  -1654.78 &    349.697 &   -339.596 &  12511.0 &   -0.130771 &   -0.445466 &   -0.477691 &   -0.397744 &    0.388782 \\
\hline
S8 & fI020 &  -1844.27 &    460.727 &   -349.145 &  10112.4 &    0.144457 &   -0.588264 &   -0.991579 &    0.514540 &    0.268374 \\
\hline
S9 & fIn024 &  -1844.27 &    460.727 &    117.911 &  10112.4 &   -0.184250 &   -0.334830 &   -2.015208 &   -0.476261 &    0.268374  \\
\hline
& Gs35\_fI020 & -1844.27 &    434.803 &   -323.221 &  10112.4 &    0.144457 &   -0.593527 &   -1.031006 &    0.514540 &    0.268374 \\
\hline
& Gs35\_fIn024 &  -1844.27 &    434.803 &    143.834 &  10112.4 &   -0.184250 &   -0.324982 &   -1.742118 &   -0.476261 &    0.268374  \\
\hline
\end{tabular}
\caption{The standard parameters of the Skyrme interactions listed in Table \ref{table1}.} \label{table2}
\end{center}
\end{table*}

\section{Conclusions}
To summarize, we have investigated, by employing a variety of effective interactions within the TDHF approach, 
the impact of several EoS ingredients %, such as incompressibility, symmetry energy, effective mass, Lane potential derivative and surface terms 
on the exit channel (fusion vs quasifission) of nuclear reactions at energies close to
the Coulomb barrier.
In particular, we build up explicit relations between the coefficients of the Skyrme interaction and relevant nuclear properties, 
such as incompressibility, symmetry energy, effective mass, Lane potential derivative and surface terms,
in some analogy with the studies of Refs. \cite{Chen:2009wv, Chen:2010qx}. We consider EoS mainly 
differing by one ingredient, with respect to a reference case, to focus on the effect of that particular ingredient on the reaction process. In such a way, we are able to decouple possible correlations among the different sectors of the EoS. The trajectory of the reaction is traced by evaluating the quadrupole moment $Q_2(t)$ of the composite system or by looking at the density contour plots. 
%on the reaction plane. \textcolor{blue}{The contour plot is not on the reaction plane now because we adopt Scamps' suggestion, it is perpendicular with the reaction plane. Maybe it is called out of plane.}
Calculations are shown for the reaction $^{238}$U+$^{40}$Ca at $E_{cm}=203$ MeV and zero impact parameter. 
We observe that all the ingredients listed above contribute to the competition between fusion and quasifission processes, however the leading role in determining the outcome of the reaction is played by incompressibility, symmetry energy and the isoscalar coefficient of the surface term. 
These results enable us to establish possible connections between the reaction dynamics and global nuclear matter properties, opening the perspective to learn on
specific aspects which are still poorly known. 
We also note that, when quasifission is observed, the features of the 
two final fragments may depend on the effective interaction considered, also
in connection with the contact time between the two interacting nuclei.  
Results concerning the corresponding charge/mass and energy 
sharing will be the object of a forthcoming publication.
We finally stress that a deeper understanding of the interplay between fusion and quasifission processes in low energy heavy-ion collisions is instrumental for the search of new SHE in the laboratory. 

\section{Acknowledgments}
This project has received funding from the European Union's Horizon 2020 research and innovation programme under grant agreement N. 654002.

\appendix
\section{}
\label{app:coeff}
%\section{
The relations between the $C..$ and $D..$ coefficients of the energy density functional of Eq.(\ref{eq:rhoE}) and the standard parameters of the Skyrme interactions are given below:
%\begin{align}
%C_0 &= \dfrac{3}{8} t_0, \nonumber \\
%D_0 &= - \frac{1}{8} t_0 \left (2 x_0 + 1 \right ),\nonumber \\
%C_3 &= \dfrac{1}{16} t_3, \nonumber \\
%D_3 &= - \frac{1}{48} t_3 \left (2 x_3 + 1 \right ),\nonumber \\
%C_{eff} &= \dfrac{1}{16} \left [ 3 t_1   + t_2  \left (4 x_2 + 5 \right ) \right ],  \nonumber \\
%D_ {eff} &= - \frac{1}{16} \left [  t_1  \left ( 2 x_1 + 1 \right ) - t_2 \left ( 2 x_2 + 1 \right ) \right ], \nonumber \\
%C_{surf} &= \dfrac{1}{64} \left [ 9 t_1  - t_2 \left (4 x_2 + 5 \right ) \right ], \nonumber \\ 
%D_{surf} &= -\frac{1}{64} \left [ 3 t_1  \left ( 2 x_1 + 1 \right ) + t_2 \left ( 2 x_2 + 1 \right )\right ] . 
%%&C_{so}= \dfrac{3}{4} W_0 &&D_{so} = \frac{1}{4} W_0\nonumber \\
%%&C_{sg}= \dfrac{1}{32} \left [t_1 \left ( 1 - 2x_1\right ) - t_2 \left ( 1 + 2x_2 \right ) \right ] \qquad &&D_{sg} = \frac{1}{32} \left ( t_1 - t_2 \right ).
%\end{align}
\begin{equation}
C_0 = \dfrac{3}{8} t_0,
\end{equation}
\begin{equation}
D_0 = - \frac{1}{8} t_0 \left (2 x_0 + 1 \right ),
\end{equation}
\begin{equation}
C_3 = \dfrac{1}{16} t_3, 
\end{equation}
\begin{equation}
D_3 = - \frac{1}{48} t_3 \left (2 x_3 + 1 \right ),
\end{equation}
\begin{equation}
C_{eff} = \dfrac{1}{16} \left [ 3 t_1   + t_2  \left (4 x_2 + 5 \right ) \right ], 
\end{equation}
\begin{equation}
D_ {eff} = - \frac{1}{16} \left [  t_1  \left ( 2 x_1 + 1 \right ) - t_2 \left ( 2 x_2 + 1 \right ) \right ], 
\end{equation}
\begin{equation}
C_{surf} = \dfrac{1}{64} \left [ 9 t_1  - t_2 \left (4 x_2 + 5 \right ) \right ], 
\end{equation}
\begin{equation}
D_{surf} = -\frac{1}{64} \left [ 3 t_1  \left ( 2 x_1 + 1 \right ) + t_2 \left ( 2 x_2 + 1 \right )\right ] . 
\end{equation}
{From the $C..,D..$ coefficients of Eq.(\ref{eq:rhoE}), the Lane potential Eq. (\ref{lanep}) and its derivative Eq. (\ref{dudp}) can be evaluated. 
%
%
%\section{Standard Skyrme parameters}
%\label{app:standard}
In Table \ref{table2} we list the parameters of the Skyrme interactions employed in 
our study, in their standard form.
%\textcolor{red}{[delete] To extract the interaction parameters, we first exploit 
%%  These parameters are determined by first considering 
%the relations between the nuclear matter properties that we want to reproduce, and the $C..,D..$ coefficients of Eq.(\ref{eq:rhoE}) \cite{Chen:2009wv, Chen:2010qx}.
%Then, inverting the relations given in Appendix \ref{app:coeff}, one can finally
%derive the standard Skyrme parameters. } \textcolor{blue}
{In analogy with the studies of Refs. \cite{Chen:2009wv, Chen:2010qx}, the standard Skyrme parameters are derived imposing to reproduce nuclear matter properties and surface effects (see Section II).  For the spin-orbit term, the coefficients corresponding to the SAMi-J
interactions are adopted for all the interactions considered in our study:
%\textcolor{red}{$W_0$ = 137 and  $W'_0$ = 42} \textcolor{blue}
{$W_{01}$ = 216.874 and  $W_{02}$ = -133.570}.}

\end{document}